\documentclass{amsart}

\usepackage{amssymb}

\newtheorem{theorem}{Theorem}[section]
\newtheorem{lemma}[theorem]{Lemma}
\newtheorem{corollary}[theorem]{Corollary}

\theoremstyle{definition}
\newtheorem{definition}[theorem]{Definition}
\newtheorem{example}[theorem]{Example}
\newtheorem{guess}[theorem] {Proposition}
\newtheorem{notation}[theorem]{Notation}
\theoremstyle{remark}
\newtheorem{remark}[theorem]{Remark}

\numberwithin{equation}{section}

\font\cal=eusb10
\def\square#1{\vbox{\hrule
\hbox{\vrule\hbox to #1 pt{\hfill}\vbox{\vskip #1 pt}\vrule}\hrule}}

\begin{document}

\title{Three applications of instanton numbers}

\author{Elizabeth Gasparim}
\address{Department of Mathematical Sciences, New Mexico State
University, Las Cruces NM 88003-8001} 
\email{gasparim@nmsu.edu}

\author{Pedro Ontaneda}
\address{Department of Mathematical Sciences,
SUNY at Binghamton, 
Binghamton, NY 13902-6000}
\email{pedro@math.binghamton.edu}

\thanks{ The authors acknowledges support from 
NSF and NSF/NMSU Advance.} 
\vspace{7mm}

\begin{abstract}{ We use  
instanton numbers to: (i) stratify moduli of vector bundles, 
(ii) calculate relative homology of moduli spaces and (iii)  
distinguish  curve singularities.  }
\end{abstract}

\maketitle

\section{Introduction} 
Instantons on a blow-up have two local numerical invariants,
which we name  {\em height} and {\em width}.
Their sum gives
the instanton charge.  In this paper we present some ways in which 
this pair of invariants gives finer information 
than the charge alone. 
Firstly, we
 show that instanton numbers give the coarsest 
stratification of moduli of bundles on blow-ups for which 
the strata are separated.
Secondly, we show that
the relative homology
$H_2({\mathfrak M}_k(\widetilde{X}), {\mathfrak M}_k(X))$ in nontrivial;
where ${\mathfrak M}_k$ denotes  moduli of charge k instantons, and 
$\widetilde{X}$ is obtained from $X$ by blowing up a point.  
This shows that, despite the fact that ${\mathfrak M}_k(X)$ 
and ${\mathfrak M}_k(\widetilde{X})$ have the same dimension, there is 
a significant topological difference between them.
Thirdly, we give examples of  analytically
distinct curve singularities, which are not distinguished by any of the 
classical invariants ($\delta_P,$ Milnor number, Tjurina number,
and multiplicity)
but have distinct 
instanton numbers. 

This paper focuses on rank 2 instantons on 
blown-up surfaces.  We are specially interested in the behavior
of instantons near an exceptional divisor. We
give an explicit construction of instantons on 
the blow-up of 
${\Bbb C}^2$ at the origin, denoted $\widetilde{{\Bbb C}^2}.$
We show that such instantons are determined by the  data 
$\Delta \colon =(j,p,t_{\infty}),$ formed by an integer $j,$ 
a polynomial $p,$ 
and a framing at infinity, that is, a holomorphic map
$t_{\infty}\colon {\Bbb C}^2-\{0\} \rightarrow
{\mbox SL}(2,{\Bbb C}).$
The  charge of $\Delta$ 
takes values between $j$ and $j^2$ 
depending on $p.$ 
However, unlike instantons on $S^4,$
whose charge is given locally by a unique invariant, called  
the multiplicity, these instantons  have two independent
local holomorphic invariants. 
These invariants do not depend on the choice of framing, and 
 can therefore be calculated
directly from the algebraic data $(j,p).$
A Macaulay2 algorithm that calculates the instanton numbers 
out of this data is available in \cite{M2}.

The connection between  holomorphic vector bundles and instantons is
made through the Kobayashi--Hitchin correspondence.
In section 2, we  use this correspondence to 
 construct  instantons on $\widetilde{{\Bbb C}^2}.$
In section 3, we use  instanton numbers
to stratify moduli of bundles on the blown-up
plane with a fixed splitting type over the exceptional divisor.
In section 4, we consider the moduli spaces ${\mathfrak M}_k(X) $
and  ${\mathfrak M}_k(\widetilde X) $  of 
rank 2 instantons on a compact surface $X$ and on 
the surface $\widetilde{X}$ = the blow up of $X$ at a point, and prove that
 $H_2({\mathfrak M}_k(\widetilde{X}), {\mathfrak M}_k(X))\neq 0.$
In section 5, we use 
instanton numbers as invariants of curve singularities. 
For curves, the trick is as follows.
Given a plane curve  $p(x,y)=0$ with singularity at the origin,
chose an  
integer $j,$  and  construct an instanton 
with data 
$(j,p).$ We then  use the
numerical invariants  of the instanton as 
analytic invariants of the curve. 

\section{Instantons on  $\widetilde{{\Bbb C}^2}$} 

Every rank 2
 instanton on $\widetilde{{\Bbb C}^2}$ is determined
by a triple $\Delta \colon = (j,p,t_{\infty}),$
where $j$ is an integer, $p$ a polynomial and $t_{\infty}$
a trivialization at infinity.
This  characterization comes from  putting together
two results: on one side, the proof due to King  \cite{KI}
 of the Kobayashi--Hitchin correspondence 
 over the noncompact surface  $\widetilde{{\Bbb C}^2}$
and, on the other side,
 the description of rank two holomorphic bundles on 
 $\widetilde{{\Bbb C}^2}$ given in \cite{BSPM}.
We review these two results.

Instantons on the blown-up plane are naturally identified with instantons on
 $\overline{\Bbb CP}^2$ framed at infinity; this is a simple consequence of the 
fact that  $\overline{\Bbb CP}^2$ is the conformal compactification of
$\widetilde{{\Bbb C}^2}.$ 
On his Ph.D. thesis, A. King \cite{KI}
 identifies the moduli space MI$(\widetilde{{\Bbb C}^2};r,k)$
of instantons on the blown-up plane of rank $r$ and charge $k,$ with
the moduli space
MI$(\overline{\Bbb CP}^2, \infty: r, k)$
of instantons on $\overline{\Bbb CP}^2,$
framed at $\infty,$ whose underlying vector bundle
has rank $r,$ and Chern classes $c_1 = 0$ and $c_2 = k.$

On the other hand, 
consider the Hirzebruch surface $\Sigma_1,$ as
the canonical complex compactification 
obtained from  $\widetilde{{\Bbb C}^2}$ by adding a line ${\ell}_{\infty}$
at infinity. 
Essentially {\em by definition}
King  identifies the moduli space
MH$(\widetilde{{\Bbb C}^2};r,k)$ of ``stable''
holomorphic bundles on $\widetilde{{\Bbb C}^2}$ 
with rank $r$ and $c_2 = k$ with the moduli space 
MH$(\Sigma_1,\ell_{\infty};r,k)$
of holomorphic bundles on $\Sigma_1$ with a trivialization 
along $\ell_{\infty}$ 
and whose underlying vector bundle has 
rank $r,$ $c_1=0$ and $c_2=k.$
King then proves the Kobayashi--Hitchin correspondence in this case, 
namely that the map $$ \mbox{MI}(\widetilde{{\Bbb C}^2};r,k) \rightarrow 
\mbox{MH}(\widetilde{{\Bbb C}^2};r,k)$$
given by taking the holomorphic part of 
an instanton connection is a bijection.
Therefore, a rank 2 instanton on 
$\widetilde{{\Bbb C}^2}$   is completely determined by 
a rank two holomorphic bundle on $\widetilde{{\Bbb C}^2}$
with vanishing first Chern class,
 together with a trivialization at infinity. 
The instanton has charge $k$ if and only if 
the corresponding holomorphic bundle 
 extends to a bundle
on $\Sigma_1$ trivial on $\ell_{\infty}$ having $c_2=k.$
We are led to study holomorphic rank two bundles on 
$\widetilde{{\Bbb C}^2}.$ 
As shown in  \cite{JA},
 holomorphic bundles on $\widetilde{{\Bbb C}^2}$
 are algebraic extensions of line bundles;
  moreover, by \cite{CA}, if the first Chern class vanishes, then 
such bundles are trivial on 
the complement of the exceptional divisor.
\vspace{3mm}

\noindent{\it Note}:
Triviality outside the exceptional divisor is 
very useful and is intrinsically related to the fact that 
holomorphic bundles on 
 $\widetilde{{\Bbb C}^2}$ are algebraic, cf. \cite{CA}.
 It is of course not true in general that
a holomorphic bundle defined only 
on $\widetilde{{\Bbb C}^2}$ minus the exceptional 
divisor is trivial; we make essential use of the fact that 
our bundles/instantons are defined over the entire $\widetilde{{\Bbb C}^2}.$
\vspace{3mm}

Now that we have established the equivalence between instantons and 
bundles, we give an explicit construction of instantons  on
$\widetilde{{\Bbb C}^2}.$ Because of  the triviality at infinity, 
it follows that we have also existence of 
instantons on any surface containing  a 
${\mathbb P}^1$ with self-intersection -1.

A holomorphic rank 2 bundle $E$ on $\widetilde{{\Bbb C}^2}$
with vanishing first Chern class splits over the exceptional divisor 
as ${\cal O}(j) \oplus {\cal O}(-j)$ for some nonnegative integer $j,$
called the {\it splitting type} of the bundle, and, in this case, $E$ is
 an algebraic extension
\begin{equation}\label{extension}0 \rightarrow {\cal O}(-j) 
\rightarrow E \rightarrow {\cal O}(j) \rightarrow 0 \end{equation}
(here by abuse of  notation we write  ${\cal O}(k)$ 
 both for the line bundle ${\cal O}_{{\Bbb P}^1}(k)$
as well as for its pull-back to  $\widetilde{{\Bbb C}^2}$).
A bundle $E$ fitting in  an exact sequence (1) 
 is determined by its extension class 
$p \in Ext^1({\cal O}(-j), {\cal O}(j)).$
We fix, once and for all, the following  coordinate charts: 
 $$\widetilde{{\Bbb C}^2} = U \cup V$$ where  
$$U = \{(z,u)\} \simeq   {\Bbb C}^2 \simeq \{(\xi,v)\} = V$$
 with 
\begin{equation}\label{coordinates}(\xi,v)=(z^{-1}, zu)\end{equation}
in $U \cap V.$ 
Then in these coordinates, the  bundle 
$E$ has a  {\em canonical} transition 
matrix of the form  
\begin{equation}\label{matrix}\left(\begin{matrix} 
z^j & p \cr 0 &  z^{-j} \end{matrix}\right)\end{equation}
from $U$ to $V,$  where 
\begin{equation}
\label{polynomial}p \colon = \sum_{i = 1}^{2j-2} \sum_{l = i-j+1}^{j-1}p_{il}z^lu^i \end{equation}
is a polynomial in  $z, \,z^{-1}$ and $ u $
 ( \cite{JA} Thm.\thinspace2.1).
Hence $E$
 is completely determined by the pair 
$(j,p).$
To have an instanton
we need also  a trivialization at infinity.
By  \cite{CA} Cor.\thinspace 4.2,
 $E$ is trivial 
outside the exceptional divisor. Therefore 
we may assign to $E$ a trivialization at infinity $t_{\infty}
\in {\mbox SL}(2, {\Bbb C}^2 -\{0\})$
thus obtaining  an instanton.
As a consequence every rank--two
 instanton  $\Delta$ on $ \widetilde{{\Bbb C}^2}$
is determined by a triple 
\begin{equation}\label{triple}\Delta:=(j,p,t_{\infty}).\end{equation}
Generically, two 
triples
$(j,p,t_{\infty})$ and 
$(j', p', t'_{\infty})$
determine the same instanton
if an only if $j'=j,$ $p' = \lambda p$
and $t'_{\infty} = A \, t_{\infty}$
where $\lambda \neq 0,$ and 
$A \in \Gamma \left({\Bbb C}^2 - \{0\},{\mbox SL}(2,{\Bbb C})\right).$ 
To define the topological charge of $\Delta$ we need to extend 
to a compact surface.
This (local)  charge is independent of the chosen compactification,
and in fact only depends on an infinitesimal neighborhood 
of the exceptional divisor. For simplicity
we take the compactification 
given by the Hirzebruch surface $\Sigma_1$
obtained by adding to
$\widetilde{{\Bbb C}^2}$  a line at infinity.
An instanton $\Delta$ on $\widetilde{{\Bbb C}^2}$ corresponds to a bundle  
$E$ on $\Sigma_1$ trivial on $\ell_{\infty},$
together with a trivialization over this line.
Let $\pi\colon \Sigma_1 \rightarrow Z$ be the map that contracts the 
$-1$ line.
     The charge of $\Delta$ is by definition
\begin{equation}c(\Delta)=c(E)\colon = c_2(E) - c_2((\pi_*E)^{\vee\vee}).
\end{equation} 
The instanton
$\Delta$ 
is generic if and only if its charge  equals its splitting type.
Moreover,
for  every $j > 1$ there are 
nongeneric instantons $(j,p,t_{\infty}),$ with charge varying 
from $j+1$ up to  $j^2$ (see theorem \ref{pams}).

\section{Local moduli spaces}
Two triples $\Delta= (j,p,t_{\infty})$
 and $\Delta'=(j',p',t'_{\infty})$  
are equivalent if they represent the
same instanton; consequently their corresponding  bundles $E$ and $E'$ over 
 $\widetilde{{\Bbb C}^2}$
are isomorphic, hence 
 must have the same splitting type, i.e. $j=j'.$ 
Consider two triples 
$(j,p,t_{\infty}) $ and $(j,p',t'_{\infty}), $
with the same  $j,$ 
and corresponding bundles
 $(E,t_{\infty})$ 
and $(E',t'_{\infty})$
over $\Sigma_1.$
An isomorphism of  framed bundles 
is a bundle isomorphism $\Phi \colon E \rightarrow E'$
such that $\Phi(t_{\infty})=t'_{\infty}.$
Two framings  $t_{\infty}$ and $ t'_{\infty}$ for the   
same underling bundle $E$ over $\Sigma_1$
differ by a holomorphic map 
$\Phi\colon  {\ell}_{\infty}\rightarrow {\mbox SL}(2,{\Bbb C})$
and,  since $\ell_{\infty}$ is compact, $\Phi$ must be constant.
Hence, projecting  $(E,t_{\infty})$ 
on the first coordinate we obtain  a fibration of 
the space of framed bundles over $\Sigma_1$ 
 over the space of bundles over $\Sigma_1$ which are trivial 
on the line at infinity, with fibre ${\mbox SL}(2,{\Bbb C}).$

\begin{equation}\label{fibration} \begin{array}{c}
{\mbox SL}(2,{\Bbb C}) \\
\downarrow \\
\left\{ framed \, rank-2 \, bundles \, over \, \Sigma_1 \right\} \\
   \downarrow \\
\left\{ rank-2 \, bundles\, over \, \Sigma_1 \, trivial \,on \, {\ell}_{\infty} \right\}.
\end{array}\end{equation}
\vspace{3mm}

We
 are thus led to  study the base space of
this fibration.
We  define 
${\cal M}_j$ to be   space
of rank two holomorphic  bundles on the $\widetilde{{\Bbb C}^2}$
with vanishing first Chern class and with splitting type $j,$ modulo
isomorphism, that is,
\begin{equation}\label{Mj}
{\cal M}_j = \left.\left\{ \begin{array}{ll} E \,\ hol. \,\, bundle \,\,  
 over \,\, \widetilde{{\Bbb C}^2}: \\
E|_{\ell} \simeq  
{\cal O} (j) \oplus {\cal O}(-j) \end{array} \right\}\right/ \sim. \end{equation}
\vspace{2mm}

Fix the splitting type $j$ and set $J=(j-1)(2j-1).$ Then the polynomial
$p$ has $J$ coefficients and we 
identify  $p$ with the $J-$tuple of complex numbers  
formed by its coefficients written in lexicographical order. We
define 
 in ${\Bbb C}^J$ 
the equivalence relation $p \sim p'$ 
if $(j,p)$ and $(j, p')$ represent isomorphic
bundles. This gives a 
set-theoretical identification
\begin{equation}\label{quotient}  {\cal M}_j
   = {\Bbb C}^J/\sim.\end{equation}
We give ${\Bbb C}^J/\sim$
the quotient topology and  ${\cal M}_j$ 
the topology induced by (\ref{quotient}). 
${\cal M}_j$ 
is generically a complex projective space     
of dimension $2j-3$ ( \cite{JA} Thm.\thinspace3.5), and
 is included in   ${\cal M}_{j+1}$ by:

\begin{guess}\label{embedding}
 The following map defines a topological embedding 
$$\begin{array}{rcl}\Phi_j: {\cal M}_j & \rightarrow &  {\cal M}_{j+1} \cr
 (j,p) & \mapsto &  (j+1,zu^2p) \end{array}. $$
\end{guess}

The proof is in section 5.
The map $\Phi_j$ takes ${\cal M}_j$ into the least generic strata 
of ${\cal M}_{j+1}.$ In fact, im$\Phi$ equals the subset of 
${\cal M}_{j+1}$ consisting of all bundles that split on the second
 formal
neighborhood of the exceptional divisor.
The complexity of the  topology of ${\cal M}_j$ increases with $j.$
 ${\cal M}_2$ is non-Hausdorff 
and by the embedding given in proposition \ref{embedding}
 this property persists in ${\cal M}_j$ for $j \geq2.$

\begin{example}\label{m2} The description of 
${\cal M}_2 $ as a quotient $ {\mathbb C}^3/\sim$ as in  $(8)$ 
gives
 $\displaystyle {\cal M}_2 \simeq {\Bbb P}^1 \cup \{A,B\}$ where points in 
the generic set ${\Bbb P}^1$ represent bundles that do not 
split on the first formal neighborhood. The point $A$  
corresponds to a bundle 
that splits on the first
formal neighborhood but not on higher neighborhoods, and $B$ corresponds
to the split bundle 
(see \cite{JA}). The topological counterpart of 
  this decomposition is  
understood by calculating the instanton charge. If $E \in {\Bbb P}^1,$ 
then $c(E)=2$ whereas $c(A)=3$ and $c(B)=4.$ 
\end{example}

The good stratification 
of ${\cal M}_2$ by topological invariants  agrees with 
the expectation we might have based on our
experience  from the case of bundles over compact surfaces.
It then appears natural to hope that topological charges stratify 
${\cal M}_j$ into Hausdorff components. This is however entirely false. 
For each $j>2$ there are non-Hausdorff subspaces of ${\cal M}_j$ where the 
topological charge remains constant.
We now define the finer instanton numbers that stratify the 
spaces ${\cal M}_j$ into Hausdorff components.

\subsection{Instanton numbers}\label{numbers}

Consider a compact  complex  surface  $X$ 
together with the blow--up 
 $\pi
\colon \widetilde{X} \rightarrow
X$ of a point $x \in X$  and denote by 
$\ell$ the 
exceptional divisor.
By the Kobayashi--Hitchin correspondence 
instantons on $\widetilde{X}$ (resp.$X$) 
 correspond to stable bundles on $\widetilde{X}$ (resp.$X$). 

Let $\widetilde{E}$
be a holomorphic bundle 
  over 
$\widetilde{X}$ satisfying
$\det \widetilde{E} \simeq {\cal O}_ {\widetilde{X}}$ 
and
$\widetilde{E}|_{\ell}  \simeq {\cal O}(j) \oplus
 {\cal O}(-j)$ with $j \geq 0.$
Set  $E = (\pi_* {\widetilde{E}})^{\vee \vee}.$
Friedman and Morgan  (\cite{FM}, p.\thinspace393)
gave the following estimate
\begin{equation}\label{fmbound}  j \leq  
c_2(\widetilde{E}) - c_2(E) \leq j^2.\end{equation}
Sharpness of these bounds was shown in   \cite{CA2}.
Since the $n$-th infinitesimal neighborhood 
of $\ell$ on $X$
is isomorphic (as a scheme) to the $n$-th infinitesimal 
neighborhood of $\ell$ in $\widetilde{{\Bbb C}^2}$
we are able to use the explicit description 
 for bundles on $\widetilde{{\Bbb C}^2},$  
given in (\ref{matrix}).
Hence  $\widetilde{E}$ is determined 
on a neighborhood $V(\ell)$
of the exceptional divisor by a pair $(j,p).$
 Define a sheaf  $Q$ by the
exact sequence,
$$
0 \rightarrow \pi_* \widetilde{E}
 \rightarrow E \rightarrow Q \rightarrow 0.
$$
Then $Q$ is supported at the point $x$
and 
 $ c_2(\pi_*\widetilde{E}) - c_2(E) = l(Q).$
An application of
Grothendieck--Riemann--Roch  gives
$$c_2( \widetilde{E}) - c_2(E) = l(Q) +
l(R^1 \pi_* \widetilde{E}).$$
Both $l(Q)$ and
$l(R^1 \pi_* \widetilde{E})$ are local analytic invariants and depend only 
on the data $(j,p)$ defining $E$ over $V(\ell).$ 
Suppose $E$ is stable on $X,$ then $\widetilde{E}$ is stable 
on $\widetilde{X}$ see \cite{FM}. Hence, if $E$ corresponds
to an instanton on $X,$ then $\widetilde{E}$ corresponds to an
instanton on $\widetilde{E}.$ This justifies the following terminology.

\begin {definition}\label{patching}{\em A holomorphic 
bundle $\widetilde{E}$ over 
$\widetilde{X}$ such that
  $\widetilde{E}\vert_{V({\ell})} \simeq  \Delta=(j,p,t_{\infty})$
and $(\pi_*\widetilde{E})^{\vee\vee} \simeq E$  
is said to be obtained by obtained by {\em holomorphic patching} 
of $\Delta$ to $E.$ If $E$ is given a frame at the point $x$, then
$E$ and $\Delta $ uniquely determine $\widetilde{E}.$}
\end{definition}

\begin {definition}{\em We set $w(\Delta)\colon = l(Q) $ and 
 $h(\Delta)\colon = h(R^1 \pi_* \widetilde{E}) ,$ and 
call them the {\em height} and the {\em width} of $\Delta.$
The {\em charge} of $\Delta$ is given by 
$$c(\Delta):= {\bf w}(\Delta)+{\bf h}(\Delta).$$}
\end{definition}

\begin{remark} \label{program} {\em The charge addition given by the  patching of 
$\Delta$ can be calculated by a Macaulay2 program
\cite{M2}. The program has as input 
 $j$ and $p$ and as outputs  ${\bf w}(\Delta)$ and ${\bf h}(\Delta).$ }
\end{remark}

The following result shows 
that instanton numbers provide good stratifications
for moduli of instantons on $\widetilde{\Bbb C}^2.$ In fact, these 
numbers give 
the coarsest stratification of 
${\cal M}_j$ for which the strata are Hausdorff. In \cite{PAMS} 
it is shown that the stratification by Chern numbers is not
fine enough to have this property. 

\begin{theorem}\label{pams}(\cite{PAMS} Thm.\thinspace 4.1)
\label{stratification} The numerical invariants $w$
and $h$ provide a  decomposition ${\cal M}_j
= \cup S_i$ 
where each $S_i$ is homeomorphic to an open subset
of a complex projective space of dimension at most $2j-3.$
For $j > 0,$ 
the lower bounds for these invariants are $(1,j-1)$ and this pair of
invariants takes
  place on the generic part of ${\cal M}_j$
which is homeomorphic to   ${\Bbb CP}^{2j-3}$
minus a closed subvariety having codimension at least 2.
The   upper bounds  are $(j(j+1)/2, j(j-1)/2)$
and this pair occurs at the single point of ${\cal M}_j$
that represents the split bundle.
\end{theorem}

Note that the ${\cal M}_j$ are labeled by splitting type,
however, we also need the loci of fixed local charge $i.$

\begin{definition}\label{N_i} {\em The {\it local moduli} 
 ${\cal N}_i$  of bundles with fixed 
local charge $i$ is
$${\cal N}_i =
\left.\left\{ \begin{array}{ll} E \,\ hol. \,\, bundle \,\,  
 over \,\, \widetilde{{\Bbb C}^2}: c_1(E)=0, c(E)=c_2^{loc}(E) =i \end{array} \right\}\right/ \sim. $$}
\end{definition}

\begin{corollary}\label{n2} ${\cal N}_0$ is just 
a point, ${\cal N}_1$ is also just a point, and 
${\cal N}_2 \simeq {\mathbb CP}^1.$
 For $i\geq2,$ 
${\cal N}_i$ has dimension $2i-3.$
\end{corollary}

\noindent {\it Proof.} Just use theorem \ref{pams}
and 
example \ref{m2}.\hfill\square{5}

\section{Topology of instanton moduli spaces}

Let $X$ be a compact complex surface.
By the Kobayashi--Hitchin correspondence (cf. \cite{LT}), we know that 
irreducible $SU(2)$ instantons of charge $k$ on $X$
are in one-to-one correspondence with rank 2 stable holomorphic bundles on 
$X$ with Chern classes $c_1=0$ and $c_2=k.$ 
Given a complex surface $Y,$ let  ${\mathfrak M}_k(Y)$  denote
the  
moduli of irreducible instantons on $Y$ with charge $k,$
or equivalently, moduli of stable bundles on $X$  
having zero first Chern class and second
Chern class $k.$

Let 
$\pi\colon \widetilde{X} \rightarrow X$ be the blow up 
of a point $x \in X.$ 
The aim of  this section is to show that there is a significant difference  
between the moduli spaces of instantons on $X$ and $\widetilde{X},$
despite the fact that their dimensions coincide. To this 
purpose we show that 
$H_2({\mathfrak M}_k(\widetilde{X}), {\mathfrak M}_k(X)) \neq 0.$
To see  ${\mathfrak M}_k(X)$
as a subspace of ${\mathfrak M}_k(\widetilde{X}),$
the polarizations on the two surfaces have to 
be chosen appropriately.
 If $L$ is an ample divisor on $X$ then 
for large $N$ the divisor $\widetilde{L} = NL-{\ell}$ 
is ample on $\widetilde{X}.$ We fix, once and for all,
the polarizations $L$ and $\widetilde{L}$
on $X$ and $\widetilde{X}$ respectively.
 From now on,   
 ${\mathfrak M}_k(Y)$ stands for moduli of rank two
 bundles on $Y$ slope stable with respect
to the fixed polarization.
 If $E$ is  $L$--stable 
 on $X,$ then $\pi^*(E)$ is $\widetilde{L}$--
stable on $\widetilde{X}.$ Therefore, the pull back map 
induces an inclusion of moduli spaces 
 ${\mathfrak M}_k(X)
\hookrightarrow {\mathfrak M}_k(\widetilde{X}).$ We proceed to 
show that for all $k\geq 1$ the relative homology  
$H_2({\mathfrak M}_k(\widetilde{X}), {\mathfrak M}_k(X))$
does not vanish.

We use holomorphic patching as defined in \ref{patching}, and to this
end we introduce framings. 

\begin{definition}{Framed bundles. \em

\begin{itemize}

\item 
Let $\pi_F \colon 
F \rightarrow Z$ be a bundle over a surface
$Z$ that is trivial over $Z_0:= Z -
Y.$ 
Given two pairs 
$f=(f_1,f_2)\colon Z_0 \rightarrow \pi_{F}^{-1}(Z_0)$
and   $g=(g_1,g_2)\colon  Z_0 \rightarrow \pi_{F}^{-1}(Z_0)$
of linearly independent  sections of $F\vert_{Z_0},$
 we say that 
$f$ is {\em equivalent} to $g$ 
if there exist a map holomorphic 
 $\phi \colon Z_0 \rightarrow SL(2,{\mathbb C})$ 
satisfying $f=\phi g$ such that $\phi$ extends to a 
holomorphic map over the entire $Z.$ A {\it frame} of $F$ 
over $Z_0$ is an equivalence class of linearly independent sections
over $Z_0.$

\item 
 A {\em framed }
bundle $\widetilde{E}^f$ on $\widetilde{X}$ 
is a pair consisting of a bundle $\pi_{\widetilde{E}}\colon 
\widetilde{E} \rightarrow 
\widetilde{X}$ together with 
a frame of $\widetilde{E}$ over $N^0:=N(\ell) - \ell.$

\item  A {\em framed } bundle $V^f$ 
on $\widetilde{{\mathbb C}^2}$ is a
pair consisting of a bundle $\pi_V\colon
V \rightarrow
\widetilde{{\mathbb C}^2}$ together with a frame of $V$ over 
$\widetilde{{\mathbb C}^2} - \ell.$

\item  A {\em framed } bundle 
$E$ on $X$ is a pair consisting of a bundle $E \rightarrow X$ 
together with a frame of $E$ 
over  $N(x)-x,$ where $N(x)$ is a small disc neighborhood  of
$x.$ We will always consider $N(x)= \pi_{\widetilde{E}}(N(\ell)).$ 
\end{itemize}}
\end{definition}

\begin{notation} {\em ${\mathfrak M}_k^f(\widetilde{X}),$ 
${\mathfrak M}_k^f(X) ,$ and ${\cal N}_i^f$ denote the framed versions of 
 ${\mathfrak M}_k(\widetilde{X}),$ 
${\mathfrak M}_k(X) ,$ and ${\cal N}_i$ respectively.}
\end{notation}

 For any $k,$ there is a stratification  of the 
moduli space of framed bundles on $\widetilde{X}$ as
$${\mathfrak M}_k^f(\widetilde{X}) \equiv  \bigcup_{i=0}^{k}
{\mathfrak M}^f_{k-i}(X) \times {\cal N}^f_i.$$ More details 
of this decomposition are given in \cite{AJ}. 
We use the notation
 $$K_i := {\mathfrak M}^f_{k-i}(X) \times {\cal N}^f_i.$$

\begin{lemma}\label{smooth} Removing the singular points of 
${\mathfrak M}^f_k(\widetilde{X})$ does not change 
homology up to dimension $k.$   That is, if $Sing $
 denotes the singularity set of 
${\mathfrak M}^f_k(X),$ then for $q <k$
$$H_q({\mathfrak M}^f_k(\widetilde{X}))= 
H_q({\mathfrak M}^f_k(\widetilde{X})- Sing)$$
\end{lemma}
\noindent {\it Proof.} By Kuranishi theory, points
 $E \in {\mathfrak M}^f_k(\widetilde{X})$ satisfying 
$H^2(\mbox{End}_0 E) = 0$ 
are smooth points. Therefore, the singularity set of 
$ {\mathfrak M}^f_k(\widetilde{X})$ is contained in 
$\Sigma_k = \{E \in{\mathfrak M}^f_k(\widetilde{X}): H^2(\mbox{End}_0E) =0\}.$
Moreover, the moduli space is defined on a neighborhood of 
a singular point by 
$\dim H^2(\mbox{End}_0 E)$ equations. In (\cite{Do} Thm.\thinspace 5.8),
 Donaldson shows that 
$ \dim H^2(\mbox{End}_0 E) \leq a+b\sqrt{k}+3k.$ Therefore, an application
of  Kirwan's 
result (\cite{Kr} Cor.\thinspace 6.4) gives 
$H_q({\mathfrak M}^f_k(\widetilde{X}))=
H_q ({\mathfrak M}^f_k(\widetilde{X})-Sing)$ for 
$q<\dim {\mathfrak M}^f_k(\widetilde{X}) 
- 2(a+b\sqrt{k}+3k) <  \dim {\mathfrak M}^f_k(\widetilde{X})
-7k= 8k-3-7k<k.$  \hfill\square{5}
\vspace{3mm}

A similar argument holds for ${\mathfrak M}^f_k(X).$ In what 
follows we work only with the smooth part of 
 ${\mathfrak M}^f_k(\widetilde{X})$ and  ${\mathfrak M}^f_k(X)$
which, by abuse of notation,  we still denote by the same symbols.

\begin{lemma}\label{leq2} For $ q\leq 2,$ 
$$H_q( {\mathfrak M}^f_k(\widetilde{X}))= H_q(K_0 \cup K_1).$$
\end{lemma}

\noindent {\it Proof.} 
The  subset of pull-back bundles
$K_0  = \{\pi^*(E), E \in {\mathfrak M}^f_k({X})\}$ is 
well known to be 
 open and dense in ${\mathfrak M}^f_k(\widetilde{X}).$
For $i\geq 1,$  the subset
 $$K_i =  \{E \in {\mathfrak M}^f_k (\widetilde{X}) : 
c_2(\pi_*E^{\vee\vee})=k-i\} ={\mathfrak M}^f_{k-i}(X) \times {\cal N}^f_{i}$$
has real codimension  at least $2i$ by   \cite{AJ}, Lemma\thinspace 6.3.
We set $S_2= \cup_{i=2}^k K_i.$ Then $S_2$ has codimension at least
$4$ in  ${\mathfrak M}^f_k(\widetilde{X}).$ 
 Consequently, 
using lemma \ref{smooth} for $q <3,$  we have isomorphisms
$$H_q({\mathfrak M}^f_k(\widetilde{X})) = H_q({\mathfrak M}^f_k(\widetilde{X})
- {\cal S}_2 ) = H_q(K_0 \cup K_1 ).$$\hfill\square{5}

\begin{lemma}\label{K1} The real codimension of $K_1$ in 
 ${\mathfrak M}^f_k(\widetilde{X})$ 
is exactly 2.
\end{lemma}

\noindent {\it Proof.} Since $K_0$ is open and dense in 
${\mathfrak M}^f_k(\widetilde{X})$  it follows that the codimension 
of  $K_1$ in 
 ${\mathfrak M}^f_k(\widetilde{X})$ equals the codimension 
of $K_1$ inside $K_0\cup K_1.$ By definition, any bundle 
$\widetilde{E} \in K_0$ is trivial around the divisor, and 
therefore satisfies 
${\bf w} (\widetilde{E}) = {\bf h} (\widetilde{E}) =0.$ On the other 
hand, any bundle $\widetilde{F} \in K_1$ satisfies 
${\bf w}(\widetilde{F}) =1$ (by Theorem \ref{pams}).
Consequently, $K_1 = \{ \widetilde{F} \in K_0 \cup K_1 :
 {\bf w} (\widetilde{F})=1\}$
is the zero locus of a single
 analytic (in fact algebraic) 
equation in $K_0 \cup K_1$; hence $K_1$ has complex codimension  one.
\hfill\square{5}

\begin{theorem}Let $k\geq 1$ and suppose $ {\mathfrak M}^f_k(X) $
is non-empty, then
$$\displaystyle H_2({\mathfrak M}^f_k(\widetilde{X}),
{\mathfrak M}^f_k(X)) \neq 0.$$
\end{theorem}

\noindent{\it Proof.} By lemma \ref{leq2} the map
\begin{equation}\label{iso} H_q(K_0 \cup K_1)
 \rightarrow H_q ({\mathfrak M}^f_k(\widetilde{X} ))\end{equation}
 is an isomorphism, for $q=0,1,2.$

The map of pairs $(K_0 \cup K_1, K_0) \rightarrow
 ({\mathfrak M}^f_k(\widetilde{X} ) ,K_0)$ induces a map between
the long exact sequences of these pairs. Using (\ref{iso}) and the
five lemma we conclude that  the map
$$H_2(K_0 \cup K_1, K_0) \rightarrow 
H_2({\mathfrak M}^f_k(\widetilde{X} ),K_0)$$
is an isomorphism.  Since $K_1$ is closed in $K_0 \cup K_1$ we have that
$$H_2(K_0 \cup K_1, K_0)=H_2(\nu (K_1), \nu (K_1)-K_1)=H_2(T\nu (K_1))$$
 (by excision),
where $\nu (K_1)$ is the normal bundle of $K_1$ in $K_0 \cup K_1,$ and $T\nu 
(K_1)$ is the Thom space
of this bundle.
By lemma \ref{K1},  $K_1$ has codimension exactly $2,$ therefore
the fiber of $  T\nu (K_1) $ has
dimension 2. Consequently (by the Thom isomorphism or duality
Theorems):
$$H_2(T\nu (K_1) )=H_0(K_1)=r {\mathbb Z}$$
where $r$ is the number of components of $K_1,$
and it follows that  $H_2(K_0 \cup K_1, K_0)=r{\mathbb Z}.$
If  $K_1$ is connected  $H_2(K_0 \cup K_1, K_0)={\mathbb Z}.$ 
Now, the theorem follows from the simple 
observation that $K_0 $ is the set of pull-back bundles,
which is isomorphic to ${\mathfrak M}_k^f(X). $ \hfill\square{5}
\vspace{3mm}

Note that in section 2 we constructed instantons on
$\widetilde{{\mathbb C}^2}$ with any prescribed charge. However, existence of 
irreducible instantons on a compact surface  follows from existence of 
the corresponding stable bundles, what bundles in general are only known 
to exist for large $c_2.$ On a surface containing a $-1$ line, however,
many  nontrivial semistable
bundles can be constructed using our holomorphic patching \ref{patching},
by patching any $\widetilde{{\mathbb C}^2}$ instanton bundle to a trivial bundle
on $X.$

\section{Curve singularities}

 Here is how to use
instanton numbers 
  to distinguish curve singularities.
Start with a curve $p\,(x,y)=0$ on
${\Bbb C}^2.$ Choose your favorite integer $j$ 
and construct an instanton on   $\widetilde{{\Bbb C}^2}$
having data $(j,p).$ Calculate the  height, width and charge of 
the instanton, use them as invariants
of the curve.
In other words, we are using the polynomial 
defining the plane curve as an extension class in 
$\mbox{Ext}^1({\cal O}(j), {\cal O}(-j)). $ This defines a bundle 
$E(j,p)$ as in (\ref{matrix}). We then regard the instanton 
numbers of this bundle 
as being associated to the curve.

Note that 
to perform the computations  we must choose a  representative 
for the curve and coordinates for the bundle. 
Here we use
 the canonical choice of coordinates for $\widetilde{{\Bbb C}^2}$
 as in section 2 and consider only either quasi-homogeneous curves, or 
else reducible curves which are  products of two quasi homogeneous curves.
For these curves there is a preferred choice of representative.
Whereas this is certainly restrictive, it is nevertheless
true that interesting results appear, given that instanton numbers 
distinguish some of these singularities which are not distinguished by 
any of the classical invariants:
the $\delta_P$ invariant,
the Milnor number, or the Tjurina number of the singularity (see table II)
and in addition the multiplicity (table III).

Taking into account that the blow-up map in our canonical 
coordinates is given by $x \mapsto u$ and $y \mapsto zu$ 
the bundle $E(j,p)$ is represented by
$$E(j,p)\colon =\left(\begin{matrix} z^j & p(u,zu) \cr 0 & 
 z^{-j} \cr\end{matrix} \right).
$$
In this paper we give a  few results to illustrate 
the behavior of the instanton numbers when applied to singularities. 
Explicit hand-made computations of these invariants for 
small values of $j$ appear in \cite{PAMS} and \cite{CA2}.
A Macaulay2 algorithm
is available to compute the invariants in the general case,
 see Remark \ref{program}.

\begin{theorem} Instanton numbers distinguish nodes (tacnodes)
from  cusps (higher order cusps).
\end{theorem}

\noindent{\bf Proof}: These singularities have  quasi--homogeneous 
representatives of the form $y^n-x^m,$ $n < m,$ 
$n$ even for nodes and tacnodes, and
 $n$ odd for cusps and higher order cusps.
 We want to show that instanton numbers detect 
 the parity of the smallest exponent. In fact, more is true, 
instanton numbers detect the multiplicity itself.

Suppose $n_1 < n_2.$ We claim that if  $j >n_2$ then  ${\bf w}(j,p_1) \neq 
{\bf w}(j,p_2).$ In fact, for $n<m$ and large enough $j$ 
the width takes the value  $${\bf w}(j,y^n-x^m)=   n(n+1)/2.$$ 
Alternatively, by vector bundle reasons we have that
 ${\bf w}(j,p_1) < {\bf w}(j,p_2).$ The second assertion is easier 
to show. The holomorphic bundle 
$E(j,p_1)$ restricts as a non-trivial extension on  
the $n_1$th formal  neighborhood  $l_{n_1 }$   
whereas  $E(j,p_2)$ splits on  $l_{n_1 }.$ 
These bundle therefore  belong to different strata of ${\cal M}_j$
and by theorem \ref{stratification} must have
distinct instanton numbers.  \hfill\square{5}

\vspace{5mm}
\noindent We consider the following classical invariants:
\begin{itemize}
\item $\delta_P = \mbox{dim}(\widetilde{\cal O}_P/{\cal O}_P) $ 
\item Milnor number $\mu = \mbox{dim}({\cal O}/ <J(P)>)$ 
\item Tjurina number $\tau = \mbox{dim}({\cal O}/<P,J(P)>)$
\end{itemize}

\vspace{5mm }
\noindent{\bf Note}: The 
first table is motivated by
 exercise 3.8 of Hartshorne \cite{HA} page 395. However,
in the statement of the problem, 
the first polynomial contains an incorrect exponent. It is  written as 
``$x^4y-y^4$''
but it should be ``$x^5y-y^4.$''
 
\vspace{5mm}
\begin{center}
\begin{tabular}{|l|l|l|l|l|l|}
\hline
 TABLE I & & & &  \multicolumn{2}{|c|}{  $j=4$}  \\     
\hline
polynomial         &     $\delta_P$ &  $\mu$ &  $\tau$ &   ${\bf w}$ &  $h$\\
\hline
\hline
$ x^5y-y^4$             &     9 &          17 &    17 &     10 &   6\\
\hline
$ x^8-x^5y^2-x^3y^2+y^4$  &    9  &         17  &   15  &     8  &   6\\
\hline
\end{tabular}
\end{center}

\begin{theorem} In some cases instanton numbers give  finer information
than the classical
invariants.
\end{theorem}

\noindent {\it Proof.} Table II gives 2 singularities that are obviously
distinct, since they have different multiplicities, but are not distinguished 
by $\delta_p,$ Milnor and Tjurina numbers. Table III shows that 
instanton numbers are the only invariants to distinguish 
the irreducible singularity $x^3-x^2y+y^3  $
from the reducible singularity $x^3-x^2y^2+y^3. $ \hfill \square{5}

\begin{center}
\begin{tabular}{|l|l|l|l|l|l|}
\hline
 TABLE II & & & &  \multicolumn{2}{|c|}{  $j=4$}  \\     
\hline
polynomial         &   $\delta_P$ & $\mu$ &  $\tau$&   ${\bf w}$&  $h$\\
\hline
\hline
 $x^2-y^7$           &     3 &             6  &  6 &      3 &    5\\
\hline
 $x^3-y^4$           &     3    &          6  &     6 &    6   &   6\\
\hline
\end{tabular}
\end{center}
\vspace{2mm}

\begin{center}
\begin{tabular}{|l|l|l|l|l|l|l|l|}
\hline
 TABLE III & & & & & \multicolumn{3}{|c|}{  $j=4$}  \\     
\hline
polynomial & mult. & $\delta_P$ & $\mu$ & $\tau$ & ${\bf w}$ & $h$ & charge \\
\hline
\hline
$x^3-x^2y+y^3$   & 3&  3 & 4 & 4 & 4 & 3 & 7\\
\hline
$x^3-x^2y^2+y^3$  & 3& 3 & 4 & 4 & 5 & 3 & 8\\
\hline
\end{tabular}
\end{center}

\vspace{2mm}

\begin{remark}{\em  The idea of using the polynomial defining a
 singularity as the
 extension class of a holomorphic bundle can be further
 generalized in several ways. For curves themselves, one can use 
other base spaces. For instance, constructing bundles on the 
total space of ${\cal  O}_{{\Bbb P}^1}(-k)$ requires very little modifications,
but give quite different results. 
One can also generalize to hypersurfaces in higher dimensions, using 
the equation of the hypersurface to define an extension of line bundles. }
\end{remark}

\section{Embedding theorem}

\noindent{\bf Proof of Proposition 3.1:}
{\small  We want to show that 
$ (j,p) \mapsto (j+1,zu^2p)  $ defines an embedding
${\cal M}_j \rightarrow {\cal M}_{j+1}.$
We first show that the map is well defined. 
Suppose 
$ \left(\begin{matrix} z^j & p \cr 0 & z^{-j} \end{matrix}\right)$ and 
$ \left(\begin{matrix} z^j & p' \cr 0 & z^{-j} \end{matrix}\right)$ 
represent isomorphic bundles. 
Then there are coordinate changes
 $ \left(\begin{matrix}a  & b   \cr c  & d \cr\end{matrix}\right)$
holomorphic in $z, \, u$ and 
$ \left(\begin{matrix}\alpha & \beta  \cr \gamma & \delta \cr\end{matrix}\right)$
  holomorphic in $z^{-1}, \, zu$
such that 
$$ \left(\begin{matrix}\alpha & \beta  \cr \gamma & \delta \cr\end{matrix}\right)
 = \left(\begin{matrix}z^j & p'  \cr 0 & z^{-j} \cr\end{matrix}\right)
 \left(\begin{matrix}a  & b   \cr c  & d \cr\end{matrix}\right)  
\left(\begin{matrix}z^{-j} & -p  \cr 0 & z^j \cr\end{matrix}\right).$$
Therefore these two bundles are isomorphic 
exactly when the system of equations 
$$ \left(\begin{matrix}\alpha & \beta  \cr \gamma & \delta \cr\end{matrix}\right)
= \left(\begin{matrix}a + z^{-j}p'c & 
z^{2j}b +z^j(p'd  -ap) - pp'c  \cr
z^{-2j}c & d - z^{-j}pc \cr\end{matrix}\right) \eqno (*)$$
can be solved by a matrix 
 $  \left(\begin{matrix}a  & b   \cr c  & d \cr\end{matrix}\right)$
 holomorphic in $z, \, u$ which makes 
$  \left(\begin{matrix}\alpha & \beta  \cr \gamma & \delta \cr\end{matrix}\right)$
 holomorphic in $z^{-1}, \, zu.$

On the other hand, the images of these two bundles are given by
transition matrices
$ \left(\begin{matrix} z^{j+1} & z\,u^2p \cr 0 & z^{-j-1} \end{matrix}\right)$ and 
$\left(\begin{matrix} z^{j+1} & z\,u^2p' \cr 0 & z^{-j-1} \end{matrix}\right),$
which  represent isomorphic bundles iff there are
coordinate changes
  $  \left(\begin{matrix}\bar a  & \bar b   \cr \bar c  & \bar d \cr\end{matrix}\right)$
holomorphic in $z, \, u$ and 
$  \left(\begin{matrix}\bar\alpha & \bar\beta  \cr \bar\gamma & 
\bar\delta \cr\end{matrix}\right)$
  holomorphic in $z^{-1}, \, zu$ 
satisfying the equality
$$ \left(\begin{matrix}\bar\alpha & \bar\beta  \cr \bar\gamma & 
\bar\delta \cr\end{matrix}\right)
 = \left(\begin{matrix}z^{j+1} & z\,u^2p'  \cr 0 & z^{-j-1} \cr\end{matrix}\right)
 \left(\begin{matrix}\bar a  & \bar b   \cr \bar c  & \bar d \cr\end{matrix}\right)  
\left(\begin{matrix}z^{-j-1} & -z\,u^2p  \cr 0 & z^{j+1} \cr\end{matrix}\right).$$
That is, the images represent isomorphic bundles if the system
$$ \left(\begin{matrix}\bar\alpha & \bar\beta  \cr \bar\gamma & 
\bar\delta \cr\end{matrix}\right)
= \left(\begin{matrix}\bar a + z^{-j } u^2p'\bar c & 
z^{2j+2}\bar b +z^{j+2}u^2(p'\bar d  -\bar a p) - z^2u^4pp'\bar c  \cr
z^{-2j-2}\bar c & \bar d - z^{-j} u^2p\bar c \cr\end{matrix}\right) \eqno (**)$$
has a solution.

 Write $x = \sum x_i u^i$ for $x \in \{a,b,c,d,
 \bar a,\bar b,\bar c,\bar d\}$
and choose
$\bar a_i = a_{i+2},$ 
$\bar b_i = b_{i+2}u^2,$
$\bar c_i = c_{i+2}u^{-2},$
$\bar d_i = d_{i+2}.$  
Then if $ \left(\begin{matrix}a  & b   \cr c  & d \cr\end{matrix}\right)$ 
solves  (*), one verifies  that 
 $ \left(\begin{matrix} \bar a  & \bar b   \cr \bar c  & \bar d \cr \end{matrix}\right)$
solves 
(**), which implies that the images  represent 
 isomorphic bundles and therefore $ \Phi_j $  is well defined.
To show that the map is injective just reverse the 
previous argument.
Continuity is obvious. 
Now we observe also that the image 
 $\Phi_j({\cal M}_j)$ is a saturated set in ${\cal M}_{j+1}$
 (meaning that if 
$ y \sim x$ and $x \in \Phi_j({\cal M}_j)$ then $y \in \Phi_j({\cal M}_j$)).
In fact, if $E \in 
 \Phi_j({\cal M}_j)$ 
then $E$ 
splits in the 2nd formal  neighborhood.
Now if $E' \sim E$ than $E'$ must also
split in the 2nd formal  neighborhood
therefore the polynomial corresponding to $E'$ is 
of the form $u^2p'$
and hence $ \Phi_j(z^{-1}p') $ gives $E'.$
Note also that 
 $\Phi_j({\cal M}_j)$ is a closed 
subset of ${\cal M}_{j+1},$ given by
the equations 
$p_{il} = 0$ for $i = 1,2$ and  $i-j+1 \leq l \leq j-1.$
Now the fact that  $\Phi_j$ is a homeomorphism over its  
image follows from the following  easy lemma.\hfill\square{5}

\begin{lemma} 
Let $X \subset Y$ be a closed subset 
and $\sim$ an equivalence relation in $Y,$ such 
that $X$ is $\sim$ saturated. 
Then the map $I :X/{\sim} \rightarrow Y/{\sim}$
induced by the inclusion is a homeomorphism over the image.
\end{lemma}

\noindent{\bf Proof:} Denote by 
$\pi_X : X \rightarrow X/{\sim} $
and $\pi_Y: Y \rightarrow Y/{\sim}$ the projections.
Let $F$ be a closed subset of $X/{\sim}.$ Then 
$\pi_X^{-1}(F)$ is closed and saturated in $X$ and therefore 
$\pi_X^{-1} (F) $ is also closed and saturated in $Y.$ 
It follows that $\pi_Y(\pi_X^{-1}(F))$  is closed 
in $Y/{\sim}.$ \hfill\square{5}}

\end{document}